\documentclass[5p,twocolumn,10pt]{elsarticle}
\usepackage{amsmath}
\usepackage{hyperref}
\usepackage{amssymb}
\usepackage{algpseudocode}
\usepackage{algorithm}
\usepackage{caption}
\usepackage{subcaption}
\usepackage{xcolor}
\usepackage{multirow}
\usepackage{tabularx}
\usepackage{booktabs}
\usepackage{float}
\usepackage{here}
\algrenewcommand\algorithmicrequire{\textbf{Input:}}
\algrenewcommand\algorithmicensure{\textbf{Output:}}
\usepackage{amsmath}
\usepackage{array}
\usepackage{hyperref}
\usepackage[toc,page]{appendix}

\usepackage{lineno}
\usepackage{hyperref}
\usepackage{enumitem}
\modulolinenumbers[5]

\usepackage{subcaption}
\usepackage{mathtools}
\usepackage{amsmath}
\usepackage{amssymb}
\usepackage{stmaryrd}
\usepackage{amsthm}
\usepackage{amsfonts}
\usepackage{dsfont}
\usepackage{graphicx}
\usepackage{upgreek}
\usepackage{bm}
\usepackage{color}
\usepackage{float}
\usepackage{tikz-cd}
\usepackage{scrextend}
\usepackage{nicefrac}
\usepackage{algorithm,algpseudocode} 
\usepackage{todonotes}
\usepackage{diagbox}
\usepackage{booktabs}
\usepackage{boldline}
\usepackage{cleveref}

\biboptions{numbers,sort&compress}

\DeclareFontFamily{OT1}{pzc}{}
\DeclareFontShape{OT1}{pzc}{m}{it}{<-> s * [1.200] pzcmi7t}{}
\DeclareMathAlphabet{\mathpzc}{OT1}{pzc}{m}{it}

\newcommand{\bx}{{\mathbf{x}}}

\newcommand{\bk}{{\mathbf{k}}}

\newcommand{\indic}{{\mathbf{1}}}

\newcommand{\vol}{{\mathsf{vol}}}

\newcommand{\SO}[1]{{\mathrm{SO}(#1)}}

\newcommand{\mycomment}[1]{}

\theoremstyle{definition}

\newcommand{\com}[1]{} 

\newcolumntype{C}[1]{>{\centering\arraybackslash}p{#1}}

\begin{document}
\baselineskip11pt

\begin{frontmatter}

\title{Deep Neural Implicit Representation of Accessibility for Multi-Axis Manufacturing}

\author{George Harabin, Amir M. Mirzendehdel and Morad Behandish}

\address{\rm
	Palo Alto Research Center (PARC),
	3333 Coyote Hill Road, Palo Alto, California 94304
	\vspace{-15.0pt}
}

\begin{abstract} 
One of the main concerns in design and process planning for multi-axis additive and subtractive manufacturing is collision avoidance between moving objects (e.g., tool assemblies) and stationary objects (e.g., a part unified with fixtures). The collision measure for various pairs of relative rigid translations and rotations between the two pointsets can be conceptualized by a compactly supported scalar field over the 6D non-Euclidean configuration space. Explicit representation and computation of this field is costly in both time and space. If we fix $O(m)$ sparsely sampled rotations (e.g., tool orientations), computation of the collision measure field as a convolution of indicator functions of the 3D pointsets over a uniform grid (i.e., voxelized geometry) of resolution $O(n^3)$ via fast Fourier transforms (FFTs) scales as in $O(mn^3 \log n)$ in time and $O(mn^3)$ in space. In this paper, we develop an implicit representation of the collision measure field via deep neural networks (DNNs). We show that our approach is able to accurately interpolate the collision measure from a sparse sampling of rotations, and can represent the collision measure field with a small memory footprint. Moreover, we show that this representation can be efficiently updated through fine-tuning to more efficiently train the network on multi-resolution data, as well as accommodate incremental changes to the geometry (such as might occur in iterative processes such as topology optimization of the part subject to CNC tool accessibility constraints).

\end{abstract}


\end{frontmatter}
\section{Introduction}
\label{sec_introduction}
Multi-axis manufacturing techniques such as multi-axis machining are widely used for manufacturing high-quality reproducible parts across multiple industries including aerospace and automotive. In multi-axis machining, one begins with a raw stock of material which is carved until the desired shape emerges. Compared with lower degree of freedom (DOF) alternatives, multi-axis machining allows for higher quality parts, greater geometric complexity, and a reduction in manual labor~\cite{davim2008machining}. Recently, multi-axis additive technologies have attracted interest, with the number of publications on multi-axis AM increasing greatly in the last 10~years. Jiang et al. \cite{jiang2021multiaxisam} show that multi-axis AM technologies may achieve greater geometric complexity, improved part quality, and reduction in support material compared to traditional 3-axis additive manufacturing. Multi-axis manufacturing affords a much larger configuration space which can help to overcome collision issues during the manufacturing process~\cite{kaji2022procplan}, however, with the rise of these technologies it will become increasingly important to incorporate this expanded freedom early in the design process.

Previous works by Mirzendehdel et al. \cite{mirzendehdel2020topology, mirzendehdel2021optimizing, mirzendehdel2022topology} have incorporated accessibility constraints for multi-axis machining into the design optimization process through the use of a continuous inaccessibility measure field (IMF), which is closely related to the concept of the Configuration Space Obstacle (CSO). This work relies on a voxelized representation of the tool or part, and is limited to considering translational motions for a fixed orientation of the tool. As a result, constructing a high-fidelity representation of the IMF for many different tool orientations (and potentially many different tools) requires storing voxel arrays for each tool at each orientation, with a space complexity of $\mathcal{O}(n_{T}n_{R}n_{G})$, where $n_{T}$ represents the number of tools, $n_{R}$ represents the number of orientations and $n_{G}$ represents the number of voxels in the IMF array. While this is tractable for lower resolution parts and tools, the computational complexity of computing the IMF scales as $\mathcal{O}(n_{T}n_{R}n_{G}\log(n_{G}))$, which can become costly as the grid size $n_{G}$, and the number of tools $n_{T}$, and orientations $n_{R}$ increases. 

Our approach is to implicitly represent the IMF as a fully connected feed-forward deep neural network (DNN). Our representation of the IMF possesses a number of advantages, such as being continuous (and differentiable) over the entire configuration space. Additionally, since only the weights of the network are stored, our representation is memory efficient (compared to storing multiple arrays of IMF values for different part-tool orientations), and evaluation of the network over a grid of size $n_{G}$ scales computationally as $\mathcal{O}(n_{G})$ compared to the convolution-based approach which scales as $\mathcal{O}(n_{G}\log(n_{G}))$. Finally, unlike conventional interpolation techniques, the weights of the DNN representation can easily be updated if the shape of the part or the tool is altered (as might happen in a design optimization process) through fine-tuning, in which the current weights of the model are used as the initialization for re-training. 

We note that our accessibility analysis assumes that all orientations are available to the tool at each location in the spatial domain (as is the case for ball-end milling); accessibility analysis for milling strategies such as flank-milling may be more complicated e.g.,~requiring position-based constraints on the orientation of the tool.

\subsection{Related Work}
\label{sec_related_work}
The configuration space obstacle (CSO) was introduced by Lozano-Perez \cite{lozano1983cspace}, who defined the CSO for two polyhedral (or polygonal) objects $\mathcal{A}$ and $\mathcal{B}$ to be the set of configurations $x$ of $\mathcal{A}$ (denoted $\mathcal{A}_{x}$) such that $\mathcal{A}_{x} \cap B \neq \emptyset$. Lozano-Perez showed that the CSO of two objects $\mathcal{A}$ and $\mathcal{B}$ in relative translation with each other could be represented through the Minkowksi sum of the objects point-sets as  $\mathcal{B} \oplus \mathcal{A}_{0}$, and gave an $\mathcal{O}(n)$ algorithm for computation of the CSO when $\mathcal{A}$ and $\mathcal{B}$ were convex n-gons. For objects with rotational components of configuration, Lozano-Perez approximated the complete CSO through translational "cross-sections" for a fixed relative orientation of obstacles. 

Computation of the CSO in the general case of arbitrary objects is computationally prohibitive, so prior work on computing the CSO has mostly focused on the case in which objects are represented as polyhedral convex regions (or unions of convex regions) or bitmaps (such as voxel arrays) and undergo translational motions relative to each other. In the case of polyhedral objects, the complexity of computing the boundary of the CSO is $\mathcal{O}(n^2)$ for convex polyhedra with n triangles and $\mathcal{O}(n^6)$ for non-convex polyhedra \cite{varadhan2004minkowski}. Due to the complexity of computing the boundary of the CSO through these methods, they are generally limited to low-dimensional configuration spaces \cite{pan2015configspace}. For the case in which objects are discretized as bitmaps or voxels, the computation of translational cross-sections of the CSO can be carried out through fast-Fourier transform (FFT) accelerated convolution, with a complexity of $\mathcal{O}(n_{G} \log(n_{G}))$, where $n_{G}$ represents the number of voxels in a grid -- irrespective of the geometric complexity of the underlying objects \cite{kavraki1993ComputationOC}. A disadvantage of this approach, however, is the memory required to store multiple arrays corresponding to different relative orientations of the objects when trying to construct an approximation to the full CSO. 

In the context of multi-axis manufacturing, prior work by Mirzendehdel et al. \cite{mirzendehdel2020topology, mirzendehdel2021optimizing, mirzendehdel2022topology} introduced the concept of an inaccessibility measure field (IMF), which is a continuous real-valued field defined on the configuration space of a tool (or tools) representing the measure of collision between the tool and a part (or other obstacles, such as fixtures) for a given configuration. The IMF is closely related to the Configuration-Space Obstacle (CSO), which is equivalent to the 0-superlevel set of the IMF. Mirzendehdel et al. \cite{mirzendehdel2020topology,mirzendehdel2022topology} have shown that multi-axis machining constraints can be directly incorporated in automated design frameworks (e.g., topology optimization) as a penalizing field. However, in all cases, the computation of IMF is performed over a sparsely sampled subset of rotational configurations $\Theta \in SO(3)$, while in reality, the tool can be rotated continuously at an arbitrary angle \footnote{Within the degrees of freedom of the robot and assuming no collision with its surroundings.}.

In the context of manufacturing planning, one is typically concerned whether or not the tool collides with a given part or obstacle, which can be posed as point membership classification (PMC) \footnote{This in contrast to the design optimization case, where we require a continuous field.}. In this case, one is interested in the boundary of the CSO to identify accessible regions for a given set of tools and their allowable orientations. As discussed by Nelaturi et al. \cite{nelaturi2019automatic} for support removal planning and Mirzendehdel et al. for build orientation optimization \cite{mirzendehdel2021optimizing}, accessibility analysis is a crucial step in manufacturing planning. Harabin et al. \cite{harabin2022hybrid} proposed a strategy for generating optimal hybrid manufacturing process plans through a combination of accessibility analysis to determine feasible manufacturing actions, and an informed search algorithm to determine the optimal manufacturing actions. However, in prior works the IMF has been limited to a discrete sampling of cross-sections corresponding to different build orientations, thus leading to a combinatorial optimization problem for maximizing the accessibility of support material, minimizing the total amount of required support material, or determining optimal manufacturing actions at each step of a process plan.

While the IMF provides a very general way of incorporating accessibility constraints into design optimization, computation depends on sampling the IMF for various configurations of the tool (e.g.~through voxelization of the configuration space), which is unsuitable for queries that may involve continuous rotational motion of a tool.  As new multi-axis technologies become prevalent, novel representations of the IMF that can efficiently describe continuous multi-axis motion will become increasingly important.

Typical shape representations in computer graphics fall under either mesh-based, voxel-based or point-cloud representations, however, for each of these representations there are trade-offs in terms of efficiency and accuracy. Due to their inherently continuous nature as well as their compact and efficient structure, numerous authors have looked into the possibility of representing 3D (and higher-dimensional) objects through the use of DNNs in the context of computer graphics. Park et al. proposed DeepSDF, a DNN representation of the signed-distance field (SDF) that directly regresses over the SDF, using an auto-decoder network architecture to represent SDF information for multiple shapes through a latent vector passed to the auto-decoder. This architecture was shown to provide accurate reconstructions of shapes while significantly reducing memory requirement when compared to competing representations (e.g. those using voxel/octree occupancy methods). Chabra et al. \cite{chabra2020deepls} proposed an extension of this method that replaces the global latent shape vector in DeepSDF with a series of latent codes representing shape information in a local neighborhood. Through this approach, the authors were able to significantly speed up training and decoding time, while significantly improving the accuracy of shape reconstruction. 

Recently, there has been a large amount of interest within the computer graphics community on representing scenes in a neural-implicit manner. Mildenhall et al. \cite{mildenhall2022nerf} proposed Neural Radiance Fields to represent scenes through fully-connected deep neural networks, with the input features being the spatial coordinate and viewing direction, and the output being the volume density and emitted radiance for that spatial location and viewing direction; these ideas have since been expanded in multiple directions to include multi-scale representations, large-scale and unbounded scene representations, and other novel-view synthesis tasks \cite{barron2021mipnerf, hedman2021bakenerf, mildenhall2022darknerf, verbin2022refnerf, reiser2023merf, jiang2022alignerf, barron2023zipnerf, barron2022mipnerf360}. While NeRFs couple all aspects of scene representation together through a volumetric neural-implicit representation, other works have focused on leveraging neural-implicit representations for accurate geometry reconstruction \cite{yariv2021volume, wang2021neus, yu2022monoSDF, niemeyer2020CVPR} and surface rendering \cite{cole2021diffsurf}.

In the context of robotic path planning DNN representations have been applied to representing the CSO: Pan et al. \cite{pan2015configspace} proposed an efficient method for construction of the CSO by training a support-vector machine to represent the boundaries of the CSO, and showed that the CSO could be efficiently and accurately represented with relatively few support vectors by using an active learning approach.

Our work builds upon these previous works by leveraging deep neural networks as regression functions on the C-space in order to provide a computational and memory-efficient representation of the CMF field (and by proxy, the CSO which is a sub-level set of the CMF). We show that our approach is able to provide accurate reconstructions of the IMF in the case where the objects, a part, and a 5-axis axisymmetric SM tool, are three-dimensional and have an associated 5-dimensional configuration space.

\subsection{Inaccessibility Measure Field}
Here, we review the concept of the inaccessibility measure field; for more details see \cite{mirzendehdel2020topology}. Let us assume that the tool assembly where $T = (H \cup K)$  can operate with all six degrees of freedom (three 
for translation and three for rotation) available for a rigid body, where $H$ and $K$ are the holder and the cutter, respectively. We also denote the design domain as $\Omega_{0}$, the part as $\Omega \subseteq \Omega_{0}$, and the substrate (and other fixtures) as $F$. 

Mathematically the \emph{configuration space} of rigid motions is represented as $\mathcal{C} = 
\mathbb{R}^3 \times SO(3)$, where $SO(3)$ refers to the group of 
$3\times 3$ orthogonal transformations that represent spatial rotations. 

In contrast to Mirzendehdel et al. \cite{mirzendehdel2020topology}, we define the  \emph{collision measure field} (CMF) over a given tool assembly $T$'s configuration space $\mathbb{R}^3 \times SO(3)$ as the
point-wise {\it minimum} of shifted convolutions for different choices of sharp points:
\begin{equation}
	f_\text{CMF}(\textbf{x}, R; O, T, K) := \min_{\textbf{k} \in K} ~\vol
	\big[ O \cap (R, \textbf{x}) (T - \textbf{k}) \big]. \label{eq_cimf}
\end{equation}
for a query point in $\textbf{x} \in \Omega_{0}$, and orientation $R \in \Theta \subseteq \SO3$ (where $\Theta$ represents the available orientations), with obstacle $O = \Omega \cup F$. 

\begin{figure*}[h]
\centering
	\includegraphics[width=.8\linewidth]{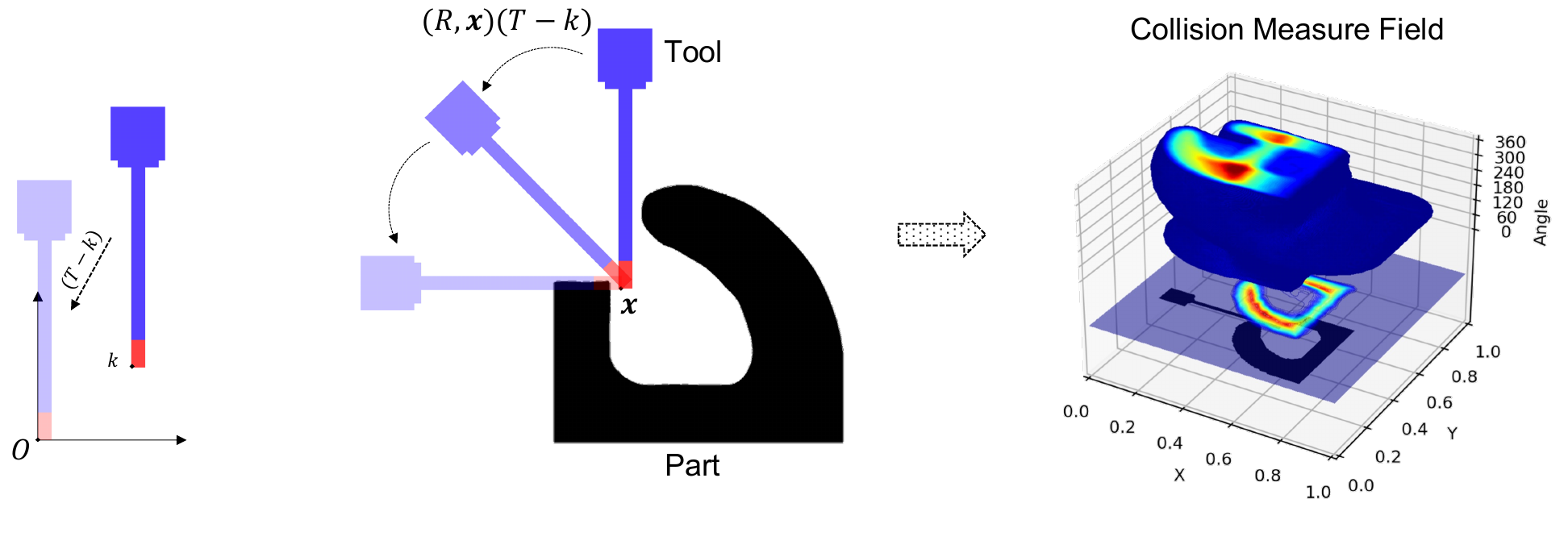}
	\caption{An illustration of the computation of the CMF and IMF for a two-dimensional tool and part. The domain of the CMF in this example is three-dimensional and can be computed for a discrete sample of orientations of the tool, while the IMF is the minimum over all such orientations.}
	\label{fig_cmf_illustration}
\end{figure*}

We also define the inaccessibility measure field over the 3D design domain $\Omega_0$ for each given tool assembly $T$ as the pointwise {\it minimum} of the CMF for different choices of available orientations $\Theta \subseteq \SO{3}$ (which depends on $T$):
\begin{equation}
	f_\text{IMF}(\textbf{x}; O, T, K) := \min_{R \in \Theta} \min_{\textbf{k} \in K} ~\vol
	\big[ O \cap (R, \textbf{x}) (T - \textbf{k}) \big]. \label{eq_imf_1}
\end{equation}
We depict the process of computing the CMF and the IMF in Figure \ref{fig_cmf_illustration}. 

As shown in \cite{mirzendehdel2020topology}, when the part and obstacle are represented as fields through their point-wise indicator functions, the IMF can be expressed through convolution as follows:
\begin{equation}
	f_\text{IMF}(\bx; O, T, K) = \min_{R \in \Theta} \min_{\bk \in K}
	~(\indic_{O} \ast \tilde{\indic}_{RT})(\bx - R\bk). \label{eq_imf_2}
\end{equation}
Where $\indic_{O}$ is the indicator function of the obstacle, and $\tilde{\indic}_{RT}$ is the indicator function of the rotated and reflected tool. 

Throughout the rest of this paper, we choose to normalize the CMF and IMF by the measure of the tool, thus restricting its range to the unit interval.

In practice, this convolution is implemented computationally through discretization of the obstacle and tool fields as voxel arrays, at which point an algorithm such as the Fast-Fourier Transform (FFT) may be used to efficiently compute the IMF on a GPU. 

\subsection{Contributions \& Outline}
\label{sec_contr_outline}
In this paper, we present a novel and general methodology for efficiently representing the CMF (and by association the IMF) through a Deep Neural Network (DNN). We then show that our DNN representation of the IMF allows for "warm-start" training, in which a pre-trained network is used as the initialization for continued training on a new data-set, and leverage this to develop a multi-resolution fine-tuning approach which uses a mix of low and higher resolution data in order to efficiently train the network while increasing representation quality. Additionally, we show that warm-start training allows us to adjust for changes in part geometry in order to avoid the necessity of retraining a network from scratch for each new part. In summary, our contributions in this paper are threefold:

\begin{itemize}
    \item We demonstrate that high-dimensional CMFs may be efficiently represented as a DNN, that may be learned through sparse data on the configuration space.
    \item We demonstrate that our DNN representation of the CMF allows us to adjust for small changes in data with minimal retraining, and we leverage this to develop a multi-resolution fine-tuning training approach which allows us to increase the quality of the DNN representation without increasing training cost.
    \item We demonstrate that our fine-tuning approach may be used to ``warm-start" a pre-trained network reducing the number of iterations required to account for slight changes in data compared to randomly initialized networks.
\end{itemize}
\section{Proposed Method} 
\label{sec_proposed_method}

\subsection{Deep Neural Network Representation of the IMF} 
\label{sec_dnn_imf_rep}
In this section, we describe the fundamental idea and methodology behind our DNN representation of the CMF (Collision Measure Field). 

While there exist many methods for interpolation and regression of functions (such as the CMF), it is necessary to balance the complexity of the representation against its flexibility: although it is possible to fit the CMF using a regression function with enough parameters, this approach may lead to overfitting and inefficient representation of the input data. Interpolation, on the other hand, provides an exact representation of the data at selected interpolation points, however, 
selection of the appropriate basis functions depends on the geometry of the domain as well as the application, and prevents a unified approach to interpolation over more complex non-Euclidean domains such as those that may arise in multi-axis manufacturing. Deep neural networks (DNNs) offer a unified approach to regression of complex functions across diverse geometric domains, and
in recent years there has been a considerable work on using neural networks
as function representations on ``low-dimensional'' domains such as $\mathbb{R}^3 \times SO(3)$ \cite{mildenhall2022nerf,park2019deepsdf,chabra2020deepls}.

Through the universal approximation property of deep neural networks \cite{cybenko1989approximationbs, lu2017nnwidth} we are assured that the CMF, being a continuous function on the configuration space, can be represented with arbitrary accuracy through a neural network; that is, given an $\epsilon >0 $, there exists a neural network $f^{\text{nn}}_{\text{C-IMF}}(\mathbf{x}, R; O, T, K)$ such that:
\begin{equation}
    \|f^{\text{nn}}_{\text{CMF}} - f_{\text{CMF}}  \| < \epsilon.
\end{equation}

Though the universal approximation property is not constructive, it provides assurance that the class of neural networks is sufficiently expressive to accurately approximate the CMF.

\begin{figure*}[h]
\centering
	\includegraphics[width=\textwidth]{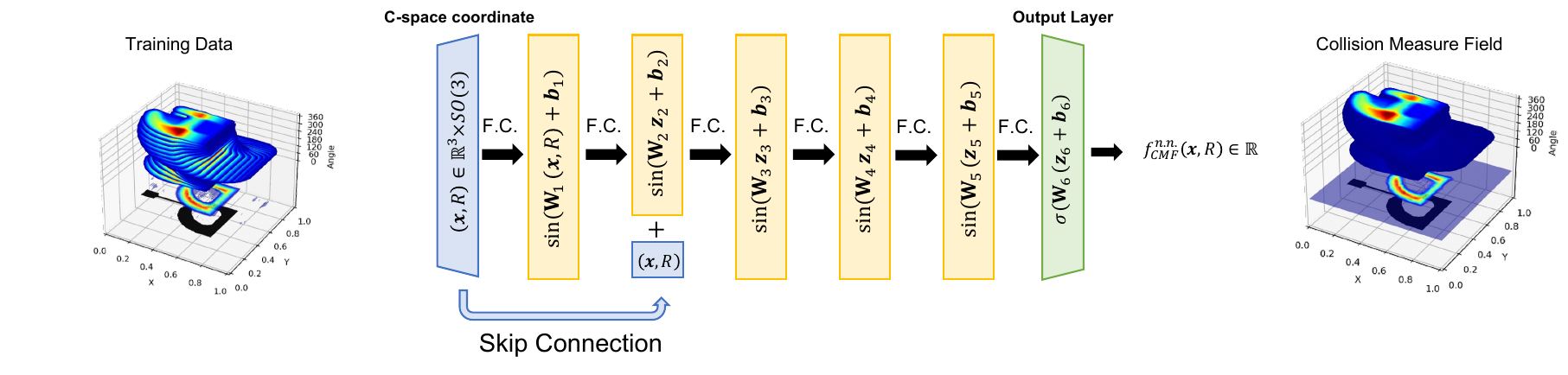}
	\caption{A diagram of the deep neural network architecture for representing the CMF and IMF described in Section \ref{sec_dnn_imf_rep}.}
	\label{fig_nn_arch}
\end{figure*}

Inspired by Mildenhall et al. \cite{mildenhall2022nerf} and Park et al. \cite{park2019deepsdf} we select a neural network architecture consisting of $d$ input features, where $d$ is the dimension of the C-space, followed by  5 fully connected layers with 512 neurons in each layer (and a skip connection that concatenates the input features to the output of the second hidden layer), and finally a sigmoid output layer that automatically constrains the output to be normalized. In light of Tancik et al. and Sitzmann et al.'s demonstration that fully connected Multi-Layer Perceptrons with ReLU non-linearities perform poorly on low-dimensional representation tasks due to spectral bias, we select a network with sinusoidal activation functions which has been shown to allow rapid learning of high-frequency features~\cite{sitzmann2020siren,tancik2020fourfeat}. A diagram of the network architecture is displayed in Figure \ref{fig_nn_arch}. 

As in \cite{park2019deepsdf}, we utilize the $L^{1}$ loss function to measure discrepancy between the network output and training samples (finding it to provide superior results to the more standard $L^{2}$ loss), and optimize the network using the Adam optimizer \cite{kingma2014adam} with weight decay $10^{-6}$, . 

In order to train the network, we pre-compute a number of voxelized cross-sections $n_{R}$ of the CMF corresponding to various orientations of the tool $\{ [f_\text{CMF}(\textbf{x}, R_{i}; O, T, K)] \mid \{ R_{i} \}_{i \in I} \subset \Theta \}$ and can be arranged in a higher-dimensional array. We interpret the values of the IMF in this array as `point-wise' values of the CMF (consisting of configuration--value tuples), and we halt the training process after a predetermined number of epochs (which are defined as one training pass through the entire data-set). 

\subsection{2D Comparison}
\begin{figure}[h]
\centering
	\includegraphics[width=.8\linewidth]{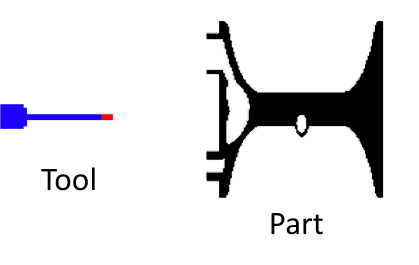}
	\caption{A depiction of the part and tool used for the two-dimensional comparison.}
	\label{fig_2D_partntool}
\end{figure}

In order to establish the suitability of our deep neural network representation of the CMF, we provide a comparison of our approach to various other interpolation schemes (trigonometric polynomials, cubic spline and linear) for a two-dimensional part and tool combination depicted in \ref{fig_2D_partntool}.

In the two-dimensional case, we assume the tool has two translational degrees of freedom corresponding to $x, y$ displacements, along with one rotational degree of freedom $\theta$. We measure IMF and CMF reconstruction quality by regressing/interpolating on $37$ input cross-sections in the interval $\theta \in [0^{\circ}, 360^{\circ}]$ (corresponding to an angular separation of $10^{\circ}$ between cross-sections), and up-sampling the CMF with $145$ equispaced $\theta$ values (corresponding to an angular separation of $2.5^{\circ}$ between cross-sections) to obtain an ``up-sampled'' IMF with quadruple the angular resolution. 

\begin{figure*}[t]
\centering
	\includegraphics[width=\textwidth]{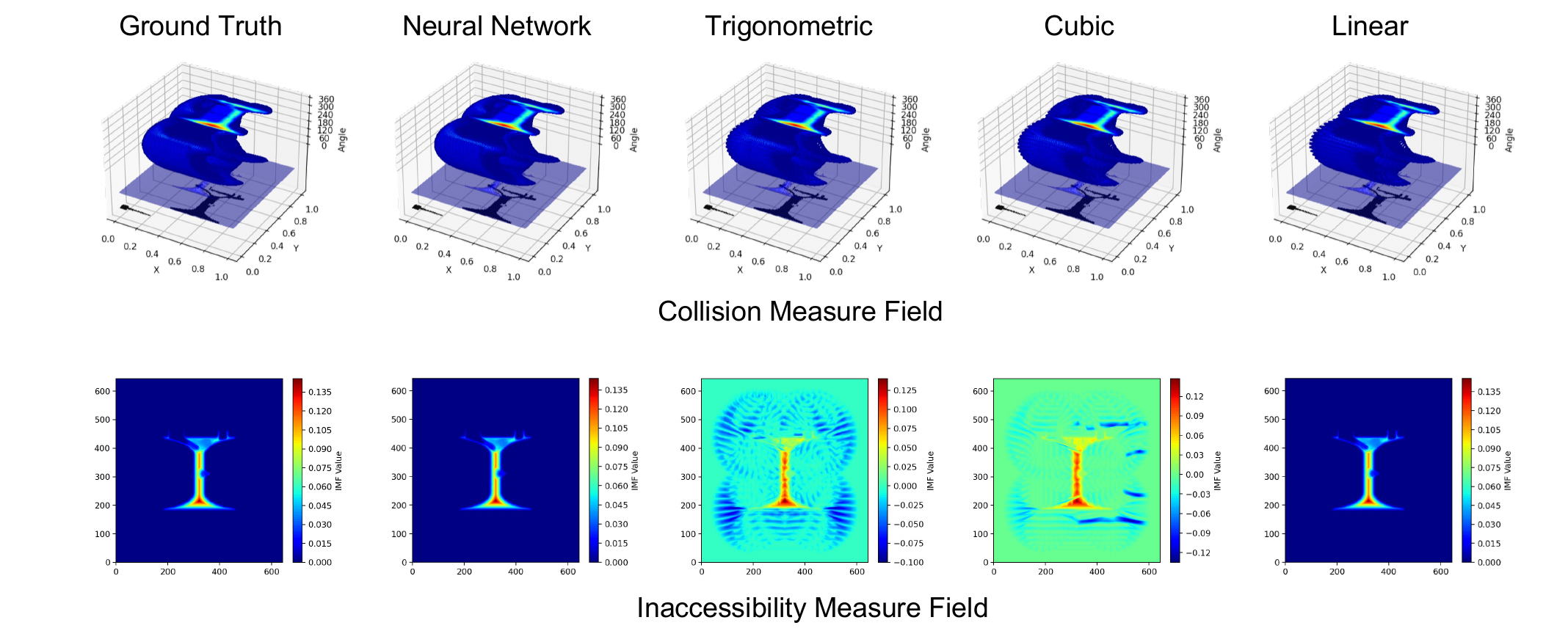}
	\caption{Comparison between our DNN representation of the CMF to other interpolation schemes.}
	\label{fig_regression_compare}
\end{figure*} 

The results are displayed in Figure \ref{fig_regression_compare}. It is visually apparent that the DNN representation of the CMF captures the correct qualitative behavior of the ground truth CMF, with each of the other interpolation methods displaying erroneous oscillations between the input cross-sections. We provide a more quantitative comparison between the IMFs produced by each method by comparing their mean-squared error (MSE) and maximum point-wise error (MPE), which is done in Table \ref{tab_2D_error}. 

\begin{table}[h!]
\caption{Error measures for different regression/interpolation methods.}
\small
\centering
\begin{tabular}{|p{1.4cm}|p{1.4cm}|p{1.4cm}|p{1.4cm}|p{1.4cm}|}
    \cline{2-5}
    \multicolumn{1}{c|}{} & DNN & Trig. & Cubic & Linear \\ \hline
    MSE & 3.7989e-7 & 2.8405e-4 & 2.3194e-4 & 5.4465e-7 \\ \hline
    MPE & 1.3659e-2 & 9.9748e-1 & 1.3398e-1 & 2.5211e-2  \\ \hline
\end{tabular}
\label{tab_2D_error}
\end{table}

Unexpectedly, the linear interpolation method outperforms the more sophisticated trigonometric and cubic interpolation methods and comes close to the quality provided by the DNN, however, this unexpected result can be explained by noting that the values of the linear interpolation between two input cross-sections must lie between the values at either end-point, and thus when taking the minimum over all rotations of the tool in Equation \ref{eq_imf_2} the interpolated values do not contribute to the computed IMF; the result shown is the same as one computed with the 37 input cross-sections. The results show that the DNN representation provides the most accurate representation of the CMF and IMF compared to the competing interpolation methods.

\section{Results}
\subsection{Single-Resolution Training}
\label{sec_sing_res}
\begin{figure*}[h]
  \centering
  \begin{subfigure}[b]{0.25\textwidth}
    \includegraphics[width=\textwidth]{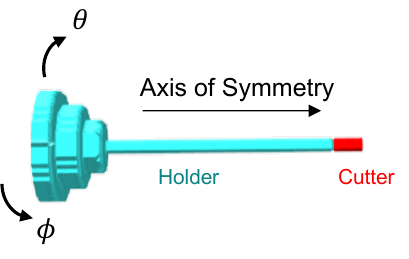}
    \caption{Tool geometry.}
    \label{fig_tool_geom}
  \end{subfigure}
  \hfill
  \begin{subfigure}[b]{0.65\textwidth}
    \includegraphics[width=\textwidth]{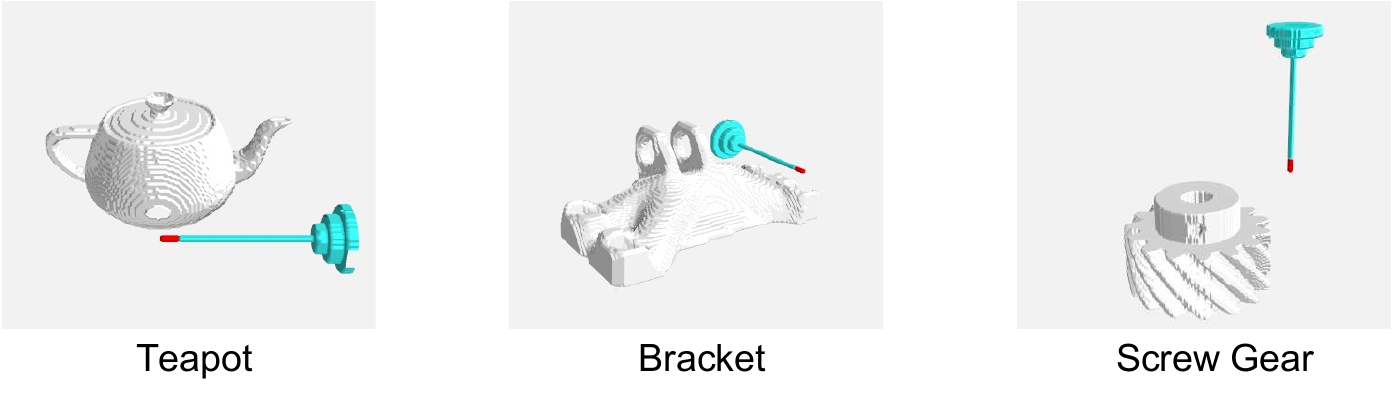}
    \caption{Part geometries used in Sec. \ref{sec_sing_res}.}
    \label{fig:subfig2}
  \end{subfigure}
  \caption{Tool and part geometries used in Sections \ref{sec_sing_res} and \ref{sec_multi_res}.}
  \label{fig_part_geom}
\end{figure*}

To demonstrate our DNN representation in the case of three-dimensional parts and tools, we analyze the accessibility of three different parts shown in Figure \ref{fig_part_geom} with respect to the tool shown in Figure \ref{fig_tool_geom}. In the case of three dimensional parts and tools, the only change to the network that is necessary is to increase the number of input features to accommodate the larger number of degrees of freedom of the tool. We choose an axisymmetric tool with a resolution of $29 \times 29 \times 83$, and discretize the teapot geometry at a resolution of $58 \times 76 \times 121$, the topology optimized bracket geometry at a resolution of $49 \times 147 \times 87$, and the screw-gear geometry at a resolution of $99 \times 98 \times 63$. Due to the axisymmetry of the tool, the rotational degrees of freedom are reduced by one, and thus the domain of the CMF is $\mathbb{R}^{3} \times S^{2}$ which we parameterize as a five-tuple $(x, y, z, \theta, \phi)$ with $\theta \in [0, 360]$ and $\phi \in [0, 180]$.

Our training data consists of 225 translational cross-sections equispaced in the angular domain, with each cross-section having the same dimensions as the discretized part, yielding over 100-million training data points for each example.  In order to reduce training time and memory footprint, we sub-sampled the training data in each cross-section with a sample density of $35 \%$, with $75 \%$ of these samples being located within the positive region of the CMF (i.e. collision) and $25 \%$ located outside of the CMF. Through this approach the network expends most of the training effort learning relevant features of the CMF rather than the non-collision portion of the configuration space. We note that before training, each point is normalized to lie within the 5-dimensional unit hypercube. We train our network for 50 epochs utilizing the Adam optimizer with an initial learning rate of $5 \times 10^{-4}$, a learning rate decay of $7 \%$ every epoch (such that the learning rate is approximately halved every 10 epochs), a weight decay $1 \times 10^{-6}$, and a batch size of 1024. All results were obtained on a Dell Precision 5820 tower with an Intel Xeon W-2275 processor, 64 GB DDR4 2933MHz RAM, and an Nvidia Quadro RTX6000 GPU.

For each example, we compare the DNN regression to the ground truth on an "in-sample" cross-section (meaning that the cross-section data was included in the training data) corresponding to a tool orientation of $\theta = 0$, $\phi = 0$, an out-of-sample cross-section corresponding to a tool orientation of $\theta = 12^{\circ}$, $\phi = 6^{\circ}$, the IMF computed with 225 cross-sections, and the IMF computed with 625 cross-sections. The in-sample cross-section and the IMF computed with 225 cross-sections provide a measure of how well the DNN can reproduce data that has been seen during training; we expect a high similarity between the DNN and the ground truth for these examples. The out-of-sample cross-section and IMF computed with 625 cross-sections provide a test of the generalization ability of the DNN since these involve data not present in the training set. 

We display our results in Figures \ref{fig_teapot}-\ref{fig_screw} and include a quantitative comparison of MSE error, and maximum pointwise error (MPE) between the DNN generated results and the ground truth in the upper half of Table \ref{tab_3D_error}. 

The teapot example provides a simple initial geometry to test our DNN representation on, with only a few thin and high-frequency features present in the handle, spout and lid. The results generated by the deep neural network are shown in Figure \ref{fig_teapot_singleres}, alongside the corresponding ground truth results displayed in Figure \ref{fig_teapot_gt}. The DNN-generated in-sample cross-section and 225 cross-section IMF show excellent qualitative and quantitative agreement with the ground truth results, with only a minor reduction in high-frequency details present on the teapot lid compared to the ground truth. Comparing the out-of sample cross-section and the IMF generated from 625 cross-sections, however, shows that the DNN representation has trouble generalizing outside of training data (with the teapot spout partially missing in the DNN generated IMF) and suggests that the angular resolution of the training data is insufficient for accurate interpolation. The quantitative comparison in Table \ref{tab_3D_error} supports the visual analysis, showing that the errors for the in-sample data are much smaller than the out-of-sample data. Surprisingly, the error for the IMF with 625 cross-sections remains relatively small (being about four orders of magnitude smaller than the out-of-sample cross-section) despite the increase in error in the out-of-sample cross-section, however, some of this difference may be explained by considering the relative magnitude of the maximum values of the CMF cross-section versus the IMF.

Next, we present the results of the DNN regression for the topology optimized bracket example in Figure \ref{fig_bracket_singleres}, along with the ground truth results in Figure \ref{fig_bracket_gt}. This example provides a greater challenge for the DNN representation due to its multiple holes and thin features. While the quantitative agreement between the in-sample cross-section and IMF generated with 225 cross-sections remains good, slight degradation can be seen around the mounting holes on the IMF, suggesting that insufficient data has been sampled from these areas. Again, the disparities observed between the DNN-generated and ground truth out-of-sample CMF cross-sections indicate that the angular resolution of the training data is insufficient for the DNN to accurately interpolate between cross-sections.

Finally, we present our results for the screw gear example in  \ref{fig_screw_singleres} with the ground truth results in  \ref{fig_screw_gt}; the screw gear has high-frequency ridges along with small features, making it a more challenging shape for regression. The DNN generated in-sample cross-section and 225 cross-section IMF show good qualitative agreement with the ground truth, moreover, there is substantially better agreement between the out-of-sample CMF cross-sections for this example. We note that compared with the teapot and bracket example, there is a much smaller difference between the in-sample and out-of-sample cross-sections for the screw gear, which may explain the network's ability to interpolate more accurately.

\subsection{Multi-Resolution Training}
\label{sec_multi_res}
The results from the previous section motivate us to investigate methods for training a more accurate DNN representation, while avoiding the computational cost of generating and training the DNN on a large number of high-resolution CMF cross-sections. In order to exploit the trade-off between training time and resolution of the training data (spatial and angular), we propose a multi-resolution ``fine-tuning'' approach in which we perform initial training of the network on a greater number of lower spatial-resolution cross-sections, ensuring sufficient angular resolution for an accurate angular regression, followed by additional training (fine-tuning) on fewer higher-resolution cross-sections in order to fine-tune the spatial resolution of the network and produce a high-quality representation of the CMF. By balancing the amount of training data in the initial training and fine-tuning stages (as well as the number of training Epochs), we ensure that the total training time of the multi-resolution approach is approximately the same as in the single-resolution training. 

We choose to test our multi-resolution training approach on the same teapot, bracket and screw-gear examples as in the previous section to provide a comparison between the single and multi-resolution approaches. For the multi-resolution examples presented, the network was initially trained on  $.65$-resolution data, with $900$ cross-sections for $20$ epochs, followed by fine-tuning the network on full-scale data with $225$ cross-sections for $25$ epochs. The same training parameters were used as in the single-resolution examples above.

\begin{figure*}
  \centering
  \begin{subfigure}[b]{\textwidth}
    \includegraphics[width=.95\textwidth]{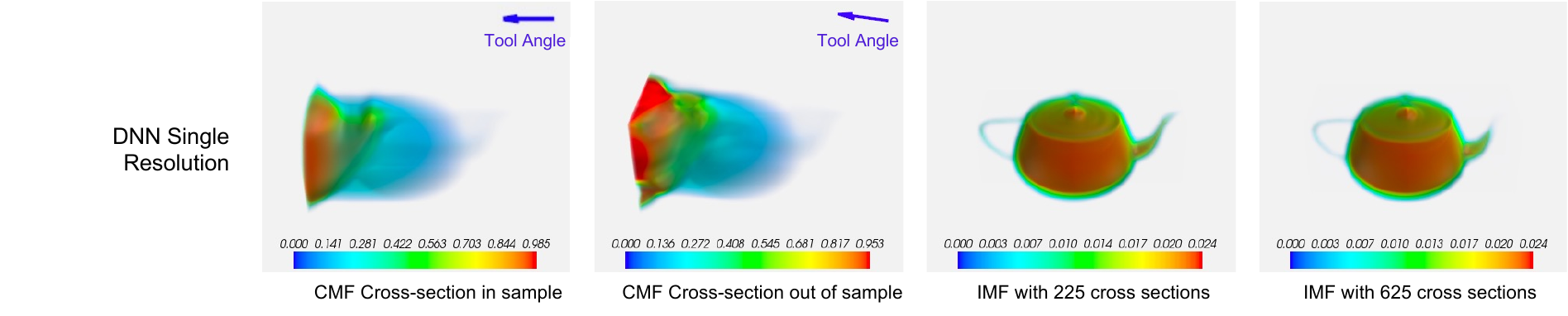}
    \caption{Single-resolution trained DNN representation of the CMF cross-sections and IMF for the Utah teapot.}
    \label{fig_teapot_singleres}
  \end{subfigure}
  \par\bigskip
  \begin{subfigure}[b]{\textwidth}
    \includegraphics[width=.95\textwidth]{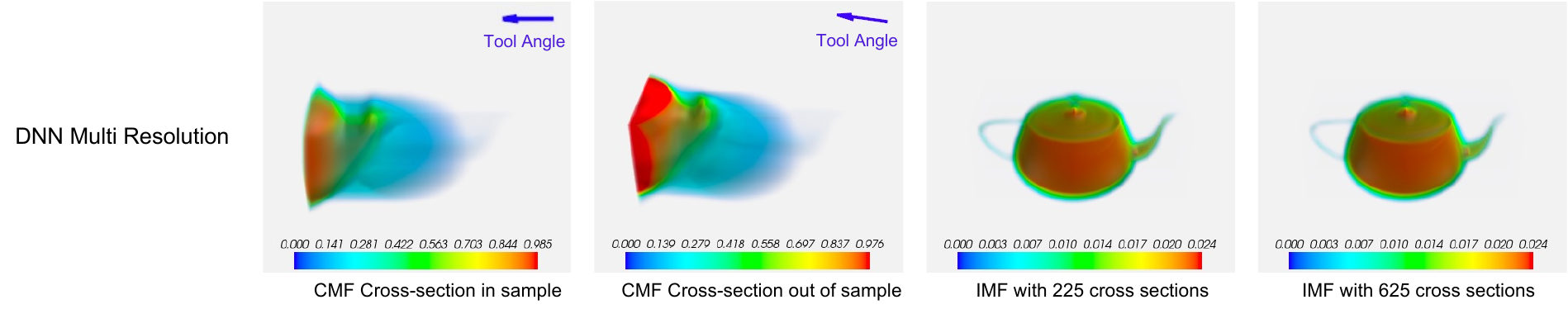}
    \caption{Multi-resolution trained DNN representation of the CMF cross-sections and IMF for the Utah teapot.}
    \label{fig_teapot_multires}
  \end{subfigure}
  \par\bigskip
  \begin{subfigure}[b]{\textwidth}
    \includegraphics[width=.95\textwidth]{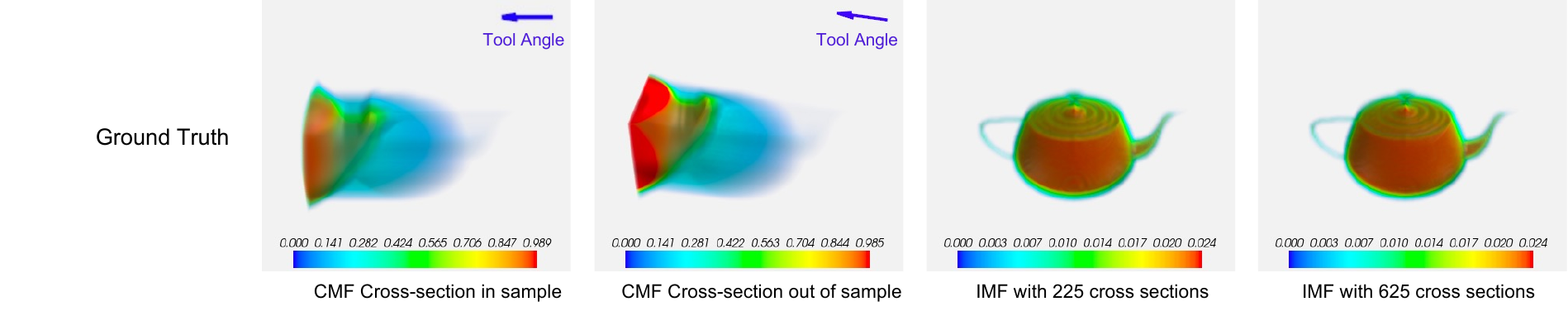}
    \caption{Ground truth CMF cross-sections and IMF Utah teapot.}
    \label{fig_teapot_gt}
  \end{subfigure}
  \caption{Ground truth and DNN regression results for the teapot geometry.}
  \label{fig_teapot}
\end{figure*}

\begin{figure*}
  \centering
  \begin{subfigure}[b]{\textwidth}
    \includegraphics[width=.95\textwidth]{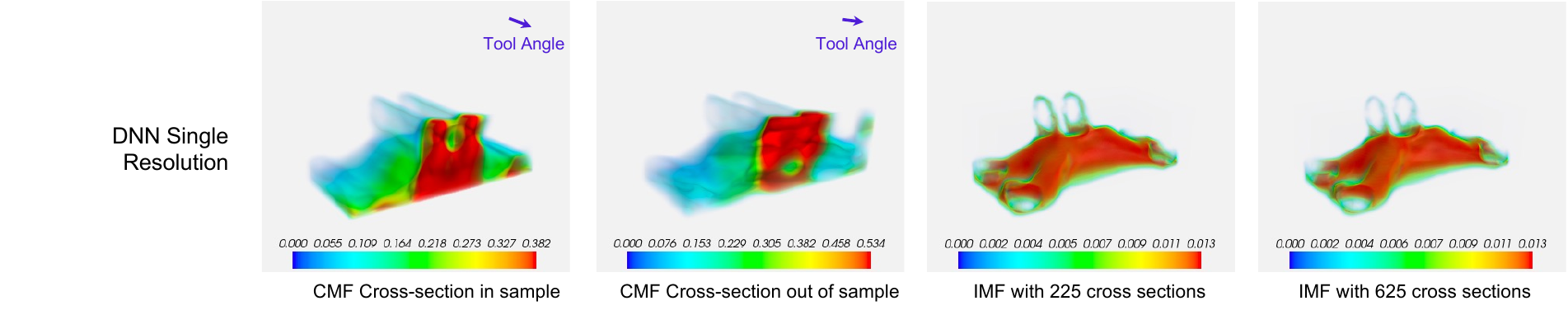}
    \caption{Single-resolution trained DNN representation of the CMF cross-sections and IMF for the topology optimized bracket example.}
    \label{fig_bracket_singleres}
  \end{subfigure}
  \par\bigskip
  \begin{subfigure}[b]{\textwidth}
    \includegraphics[width=.95\textwidth]{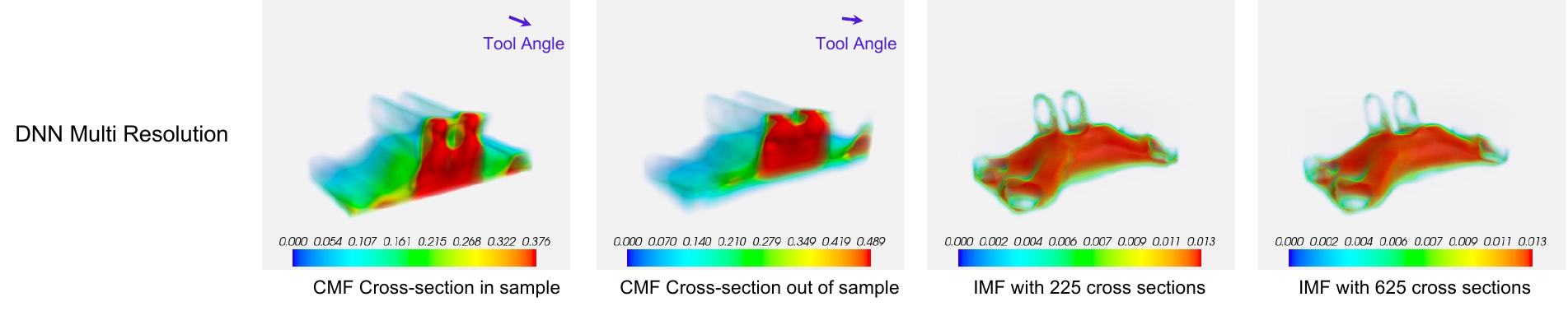}
    \caption{Multi-resolution trained DNN representation of the CMF cross-sections and IMF for the topology optimized bracket example.}
    \label{fig_bracket_multires}
  \end{subfigure}
  \par\bigskip
  \begin{subfigure}[b]{\textwidth}
    \includegraphics[width=.95\textwidth]{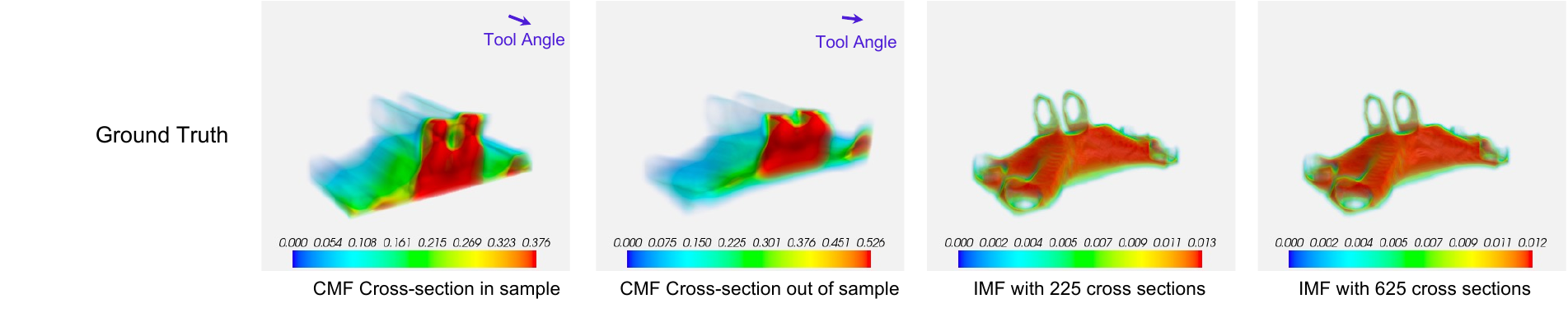}
    \caption{Ground truth CMF cross-sections and IMF for the topology optimized bracket example.}
    \label{fig_bracket_gt}
  \end{subfigure}
  \caption{Ground truth and DNN regression results for the bracket geometry.}
  \label{fig_bracket}
\end{figure*}

\begin{figure*}
  \centering
  \begin{subfigure}[b]{\textwidth}
    \includegraphics[width=.95\textwidth]{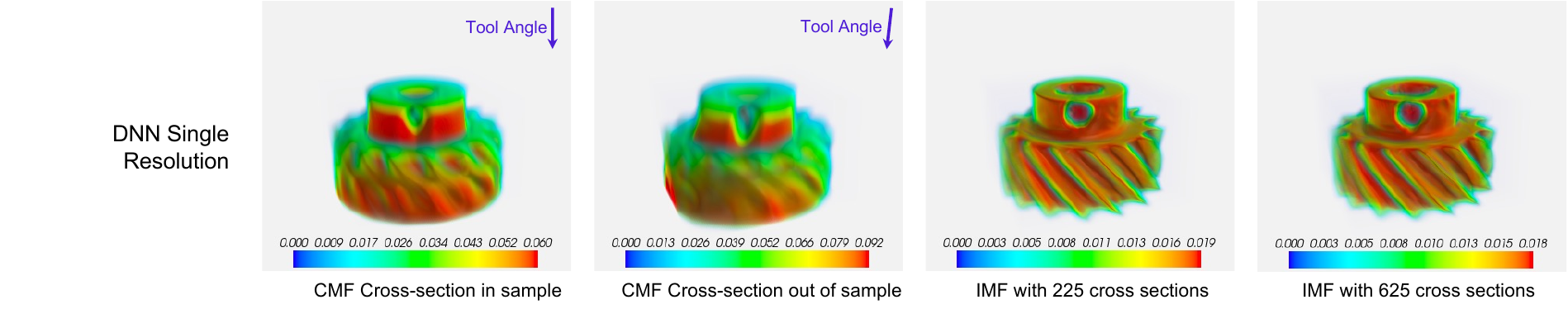}
    \caption{Single-resolution trained DNN representation of the CMF cross-sections and IMF for the screw gear example.}
    \label{fig_screw_singleres}
  \end{subfigure}
  \par\bigskip
  \begin{subfigure}[b]{\textwidth}
    \includegraphics[width=.95\textwidth]{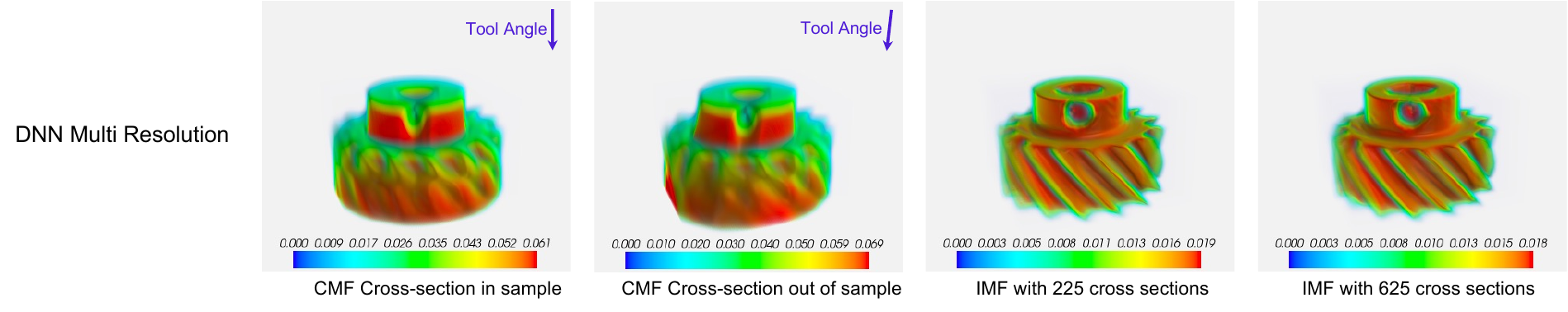}
    \caption{Multi-resolution trained DNN representation of the CMF cross-sections and IMF for the screw gear example.}
    \label{fig_screw_multires}
  \end{subfigure}
  \par\bigskip
  \begin{subfigure}[b]{\textwidth}
    \includegraphics[width=.95\textwidth]{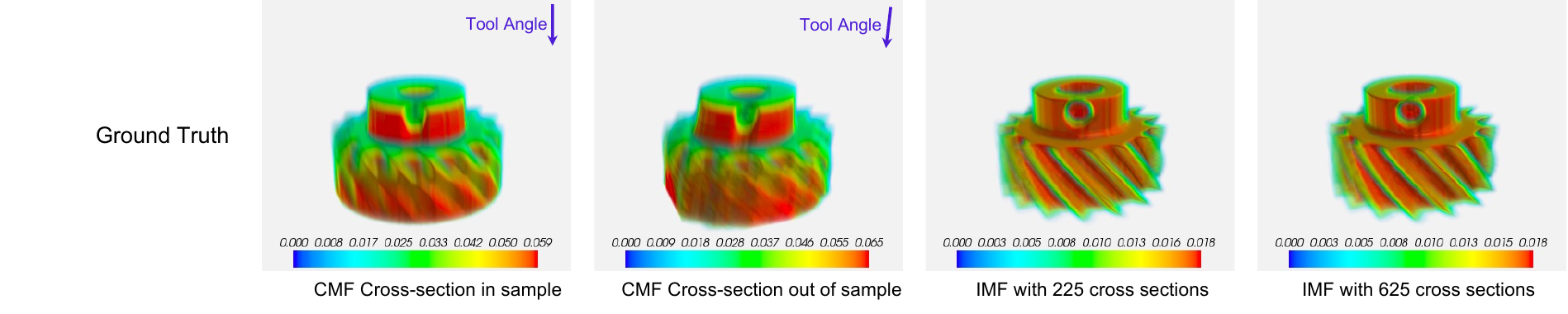}
    \caption{Ground truth CMF cross-sections and IMF for the screw gear example.}
    \label{fig_screw_gt}
  \end{subfigure}
  \caption{Ground truth and DNN regression results for the screw gear geometry.}
  \label{fig_screw}
\end{figure*}

Our results for the multi-resolution training are displayed in Figures \ref{fig_teapot_multires}, \ref{fig_bracket_multires} and \ref{fig_screw_multires}. We also provide a quantitative comparison to the ground truth in Table \ref{tab_3D_error} under the Multi-Res. row. 

Visual inspection of the second rows of  \cref{fig_teapot,fig_screw} reveals that the multi-resolution out-of-sample CMF cross-sections, and IMF with 625 cross-sections are more accurate compared to the single resolution examples, while the quantitative error measurement shows that the multi-resolution training process does not adversely affect the accuracy of the in-sample cross-section and 225 cross-section IMF. Thus the multi-resolution training allows the neural network to gain the advantages of higher angular-resolution training data during the initial training, along with the higher spatial-resolution data from the fine-tuning. 

There are, however, still some visual and quantitative discrepancies between the multi-resolution and the ground truth results: most notably, the out-of-sample CMF cross-sections for the multi-resolution trained DNN suffer from quantitative discrepancies compared to the ground truth (e.g.~the maximum value of the CMF) in the bracket example, as well as lack of detail in small (high frequency) features apparent in the screw gear example. 

\begin{table*}
\caption{Error measures for the DNN regression of the CMF cross-sections and the IMF.}
\centering
\small
\begin{tabular}{|c|c|l|C{3cm}|C{3cm}|C{3cm}|C{3cm}|}
\cline{4-7}
\multicolumn{3}{c|}{}  & CMF cross-section in-sample & CMF cross-section out-of-sample & IMF with 225 cross-sections & IMF with 625 cross-sections \\ \hline
\multirow{6}{*}{\rotatebox[origin=c]{90}{Single-Res.}} 
& \multirow{2}{*}{Teapot} & MSE & 7.353e-7  & .0063 & 8.188e-9 & 2.544e-7 \\ 
& & MPE& 0.0176 & 0.7009 & 0.0019 & 0.0042 \\ \cline{2-7} 
& \multirow{2}{*}{Bracket} & MSE & 6.803e-7 & 0.0004 & 1.273e-9 & 8.3764e-8 \\ 
& & MPE& 0.0207 & 0.3713 & 0.0025 & 0.0030 \\ \cline{2-7} 
& \multirow{2}{*}{Screw Gear} & MSE & 5.632e-7 & 4.456e-6
 & 5.344e-8 & 1.542e-7 \\ 
& & MPE& 0.0167 & 0.0461 & 0.0031 & 0.0035 \\ \hlineB{3}
\multirow{6}{*}{\rotatebox[origin=c]{90}{Multi-Res.}} 
& \multirow{2}{*}{Teapot} & MSE & 9.127e-7 & .0007 & 8.125e-9 & 1.794e-7 \\ 
& & MPE& 0.0151 & 0.2496 & 0.0019 & 0.0032 \\ \cline{2-7} 
& \multirow{2}{*}{Bracket} & MSE & 5.968e-7 & 2.351e-5 & 1.076e-9 & 5.6624e-8 \\ 
& & MPE& 0.0160 & 0.1017 & 0.0024 & 0.0033 \\ \cline{2-7} 
& \multirow{2}{*}{Screw Gear} & MSE & 3.871e-7 & 1.248e-5 & 4.896e-8 & 1.184e-7 \\ 
& & MPE& 0.0142 & 0.0242 & 0.0036 & 0.0034 \\ \hline
\end{tabular}
\label{tab_3D_error}
\end{table*}

The training times for each of the single-resolution examples are listed in 
\ref{tab_train_time} with each of the examples requiring a training time ranging from $10$ to $15$ hours (the multi-resolution training times were not significantly different by construction, and are not reported). Although the training time for the DNN is substantial, there are opportunities for a significant reduction in training time by employing transfer learning, as described in \ref{sec_transfer_learn}. 

\begin{table}
\caption{Training and evaluation time for DNN and convolution based approaches. All times are in seconds.}
\small
\centering
\begin{tabular}{|p{2.6cm}|p{1.4cm}|p{1.4cm}|p{1.7cm}|}
    \cline{2-4}
    \multicolumn{1}{c|}{} & Teapot & Bracket & Screw Gear \\ \hline
    DNN Train Time & 36105 sec. & 40760 sec. & 56235 sec. \\ \hline
    DNN Eval. Time & 15.5 sec. & 18.0 sec. & 17.82 sec. \\ \hline
    Conv. Time & 7.0 sec. & 9.3 sec. & 7.45 sec. \\ \hline
\end{tabular}
\label{tab_train_time}
\end{table}

We also provide evaluation times for computing $225$ cross-sections of the CMF for both the DNN and the convolution-based approach (averaged over $3$ evaluations) in Table \ref{tab_train_time}. While our current network architecture yields slower evaluation times compared to the convolution-based method, our DNN representation of the CMF requires $\mathcal{O}(n_{R} n_{G})$ operations to compute $n_{R}$ cross-sections of the CMF (corresponding to one forward pass of the network for each grid point). By contrast the time complexity of computing the IMF through a the convolution-based approach for one tool assembly scales as $\mathcal{O}( n_{R} n_{G} \log(n_{G}))$ \cite{mirzendehdel2020topology} where $n_{R}$ is the number of sampled rotations of the tool assembly, and $n_{G}$ is the number of grid points. Thus, for large part and tool grids, there is an opportunity for the DNN based representation to provide computational advantage over the traditional convolution based approach. 

Finally, the memory footprint of the voxelized representation of the CMF scales in a cubic manner with the resolution of the part and tool (as $\mathcal{O}( n_{R} \cdot n_{G})$): one translational cross-section of the CMF for the teapot requires storing approximately $500,000$ floating point numbers. By comparison our DNN has roughly the same number of trainable parameters, taking roughly 2 Megabytes of storage to store the entire description of the CMF. 

\subsection{Transfer Learning}
\label{sec_transfer_learn}
While the above results demonstrate the potential of using a deep neural network to represent functions over configuration-space such as the CMF, due to the training time, it would not be feasible to train a separate neural network from scratch for each different part and tool combination. Transfer learning is a commonly used technique in machine learning in which a DNN pre-trained for one task is used as the initialization for training a model on a separate but related task \cite{ash2020warmstarting}. In this section, we demonstrate that a DNN trained to represent the CMF for one part may be efficiently retrained to account for small to moderate changes in the part geometry.

In our scenario, the weights of the DNN representation are optimized to represent the CMF of a specific part and tool combination, however, for part or tool geometries that are "close" to the initial training data, the weights of the DNN may only need slight adjustments in order to represent the new shape. The idea of choosing good initialization parameters in order to reduce training time has been explored in meta-learning algorithms such as MAML and REPTILE \cite{finn2017MAML, nichol2018reptile}, and has been exploited in the context of various DNN regression tasks by \cite{tancik2020meta}, although in this context, the initialization parameters come from optimization over multiple training examples. We provide a simple demonstration to show the potential of transfer learning to reduce the training time for a given part and tool combination from hours to minutes. 

\begin{figure*}
\centering
    \includegraphics[width=.9\textwidth]{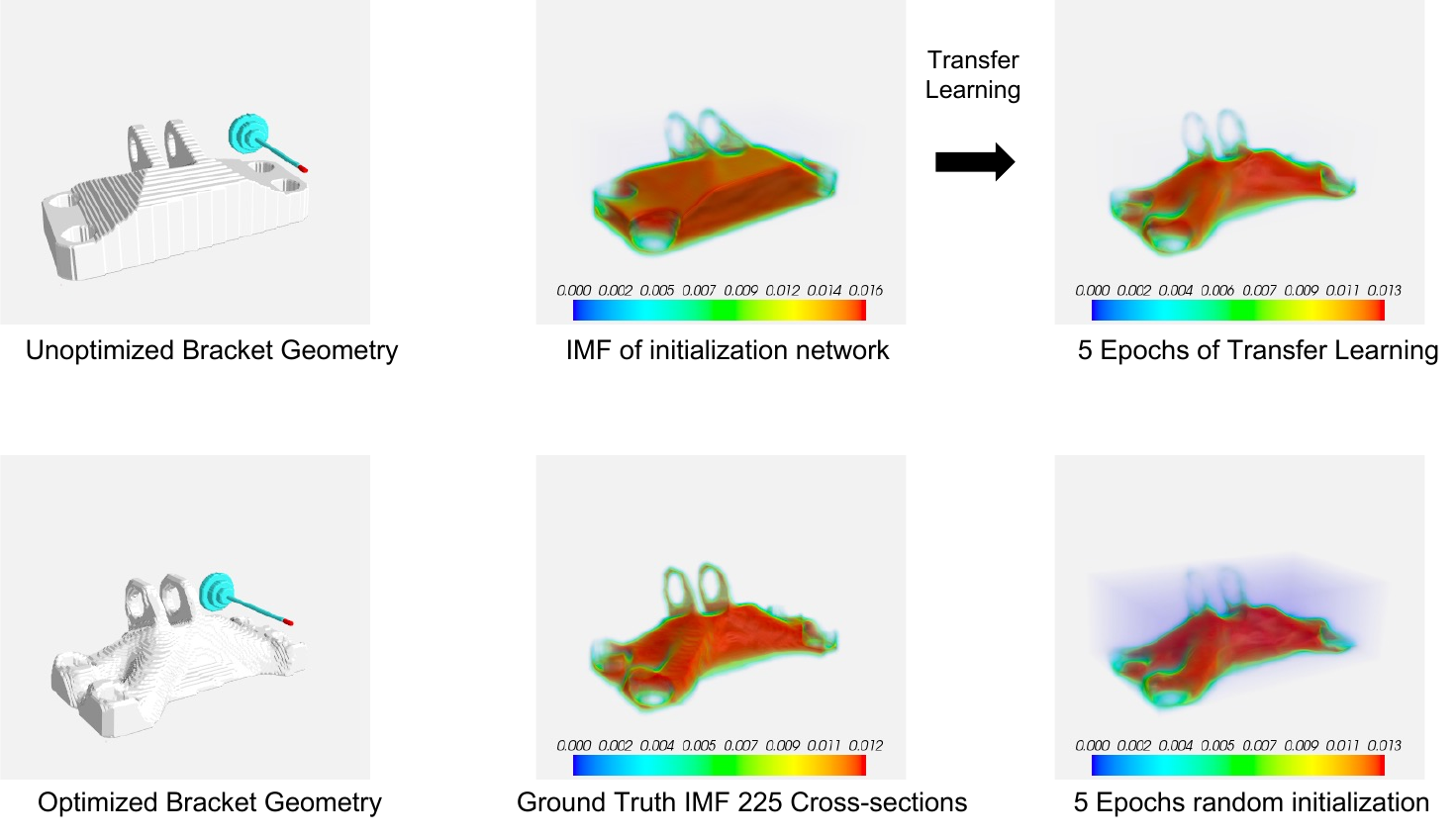}
    \caption{Transfer learning allows a pre-initialized network to be efficiently retrained to accommodate changes in part geometry.}
    \label{fig_transfer_learning}
\end{figure*}

We demonstrate our transfer-learning approach with an unoptimized bracket (depicted in  \ref{fig_transfer_learning}), which we use to pre-train the DNN that will be used as an initialization for transfer learning, and the optimized bracket (from the previous section), which we use to fine-tune the pre-trained network. We use the same training parameters and approach as in \ref{sec_sing_res} for training the initialization network, however, we only sample 15\% of the CMF cross-section data and only train for 30 epochs to demonstrate the ability of the network to deal with sparse data. This pre-training process took approximately 3~h and 17~min to complete (roughly 6.5~min per Epoch). We then fine-tune the initialization network for 5 epochs on 225 cross-sections of the optimized bracket, using the same training parameters, but starting with a reduced learning rate of $1 \times 10^{-4}$. This transfer-learning training process took approximately 29~min to complete. 

We compare the IMF produced by the DNN trained through the transfer-learning process to that of a randomly initialized DNN trained for 5 Epochs (using the same parameters). Our results are depicted in  \ref{fig_transfer_learning}, and show that the transfer-learning approach yields a much better approximation of the ground-truth IMF compared to the randomly initialized network. More quantitatively, we compare the MSE and MPE of the IMF generated through 225 cross-sections of the fine-tuned DNN vs. the randomly initialized DNN. The MSE and MPE for the fine-tuned DNN are $2.8595\text{e-}8$ and $0.0028$ respectively, while for the randomly initialized network, they are $1\text{e-}7$ and $0.0036$.

\section{Conclusion}

We have presented a novel representation of the collision measure field  as a deep neural network, and by association the inaccessibility measure field  and the configuration-space obstacle. We have shown that the collision measure field may be represented through a deep neural network with periodic activation functions, and have studied the quality of the DNN representation for various part geometries, finding that even with sparse data the CMF may be accurately reconstructed. We have also introduced a multi-resolution fine-tuning approach for improving the accuracy of the representation without increasing the training or evaluation cost of the network. Finally, we have demonstrated that transfer learning may be effectively used to reduce re-training time for similar part geometries.

As mentioned in Section \ref{sec_related_work}, the IMF has been used in topology optimization in order to enforce accessibility constraints \cite{mirzendehdel2020topology}. In this setting, the initial design is  updated in an iterative manner and the IMF, which is used as a penalizing field in order to enforce manufacturability constraints, must be re-computed at each iteration. The process of recomputing the IMF at each design iteration can become very costly in time and space if the number of design iterations and CMF cross-sections is large. In such design applications, it may be acceptable to utilize an approximate representation of the IMF such
as our DNN representation. For high-resolution parts and tools, the DNN-based
representation should scale more favorably with the spatial resolution. Additionally, recent research from Müller et al. \cite{mueller2022instant} has shown that the use of multi-resolution hash encoding may speed up the training and evaluation of neural-implicit representations by several orders of magnitude. Finally, after initial training, transfer learning enables good
approximation of the CMF with a vastly reduced training time, thus allowing the
initial training cost to be amortized over the number of optimization
iterations. These factors may enable the DNN representation to provide a computational advantage over the convolution-based approach for such applications.

While design optimization provides one potential application of our approach, the DNN representation generalizes readily to other functions whose domain is the configuration space (e.g., potential fields in path planning algorithms). Additionally, the continuity of the DNN representation permits applications to areas such as process planning for multi-axis machines, which exhibit continuous rotational motion that cannot be accurately represented through voxelized representations of the accessibility field. 

There are, however, a few limitations of our current approach which may form the basis for future work. Firstly, the training time for our DNN representation is considerable depending on the resolution of the part and tool representations. While transfer learning may be used to reduce the training time significantly for similar part and tool geometries, meta-learning approaches  may provide more efficient initialization parameters across a wider class of part and tool geometries.

Secondly, in use-cases such as process planning, where an accurate representation of the IMF is required, a principled understanding between data sparsity and regression quality is important. In this paper we have explored one data acquisition strategy (random sub-sampling of equispaced data), however, it is likely that more intelligent sampling methods (e.g. more dense sampling towards the boundary) may prove useful in minimizing the amount of data required to accurately reconstruct the CMF \cite{nelaturi2013solving}. 

While our approach provides a novel way of representing fields over configuration spaces, it is data intensive and does not leverage information about either the part or tool geometries; data-reduced and data-free methods may provide fruitful avenues for future exploration of such neural-implicit representations.

\section*{Acknowledgments}

This research was developed with funding from the Xerox Corporation. The views, opinions and/or findings expressed are those of the authors and should not be interpreted as representing the official views or policies of the Xerox Corporation.

\clearpage

\section*{References}

\appendix
\section{Data Sparsity and Architecture}
\label{sec_appendix}

In this section, we present results related to comparing different network architectures, as well as the effects of data sparsity upon regression results. We show that while our selected architecture is able to achieve good quality regression results even with sparse data, changing the network architecture dramatically affects the quality of the results. All neural networks in this section were trained with the same training parameters as in Section \ref{sec_sing_res}.

\begin{figure*}[t!]
\centering
    \includegraphics[width=.9\textwidth]{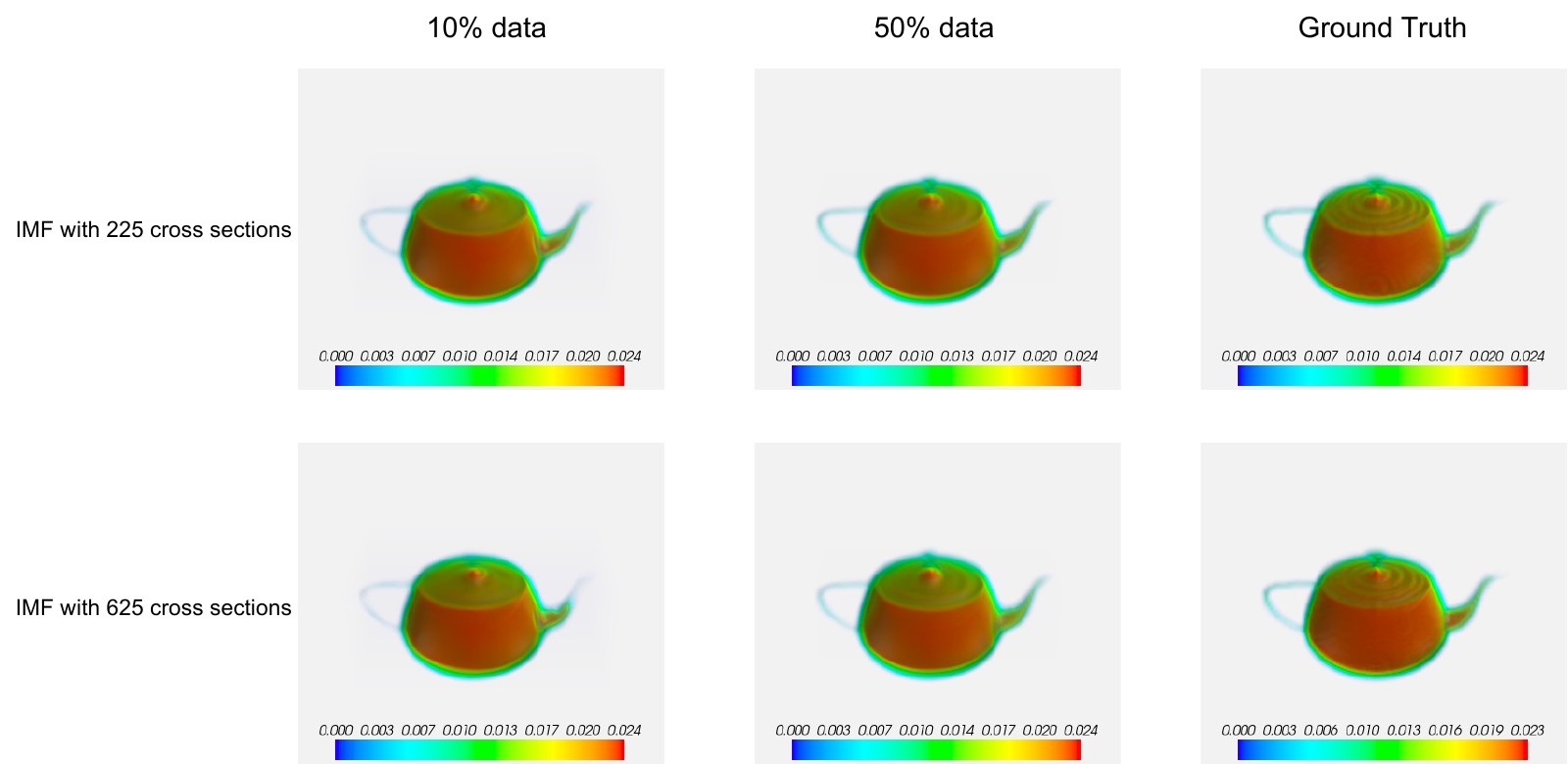}
    \caption{Effect of data sparsity upon regression results.}
    \label{fig_data_sparsity}
\end{figure*}

We begin by analyzing the effects of data sparsity upon regression quality. For this comparison, we have trained the DNN with two different sub-sample proportions: 10\% and 50\%. We analyze the effects upon the IMF generated from 225 cross-sections (in-sample data) and the IMF generated from 625 cross-sections (out-of-sample data). The results of the regression are displayed in Fig. \ref{fig_data_sparsity}, and we present a quantitative comparison in in Table \ref{tab_data_sparsity}.

\begin{table}[h!]
\caption{Error measures for different sub-sample proportions.}
\small
\centering
\begin{tabular}{|C{1.5cm}|c|C{1.5cm}|C{1.5cm}|}
\cline{3-4}
\multicolumn{1}{c}{} & & 10\% data & 50\% data \\ \hline
\multirow{2}{*}{\begin{tabular}[c]{@{}c@{}}225 \\ Cross Sec.\end{tabular}}
& MSE & 0.0019 & 0.0016  \\ 
& MPE & 1.641e-8 & 2.347e-8  \\ \hline
\multirow{2}{*}{\begin{tabular}[c]{@{}c@{}}625 \\ Cross Sec.\end{tabular}}
& MSE & 0.0039 & 0.0037   \\ 
& MPE & 8.042e-9 & 2.872e-7   \\ \hline
\end{tabular}
\label{tab_data_sparsity}
\end{table}

It is apparent from this study that the results of data sparsity upon the regression result are minimal, and are visually limited mostly to the area near the spout in the IMF with 625 cross sections.

\begin{figure*}[b!]
\centering
    \includegraphics[width=.9\textwidth]{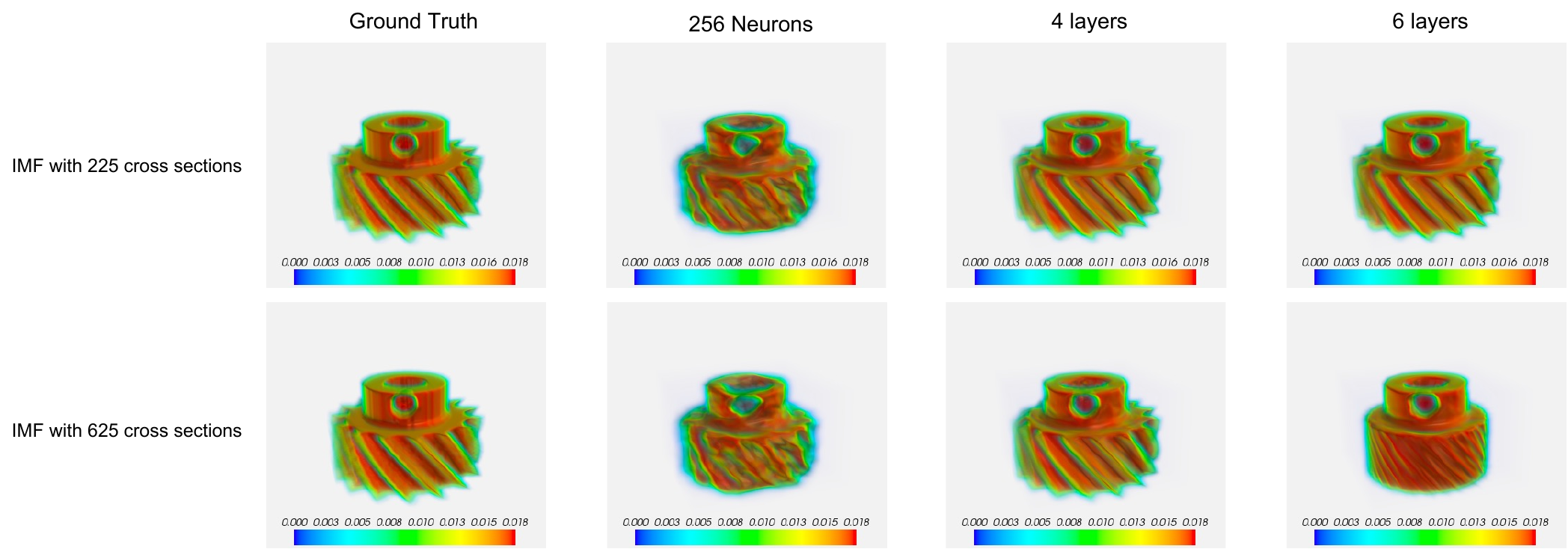}
    \caption{Effect of network architecture upon regression results.}
    \label{fig_net_arch}
\end{figure*}

We then analyze the effect of different network architectures upon the regression results for the screw gear geometry depicted in \ref{fig_part_geom}. For this test, we vary the depth and the width of the network, testing a network with half the neurons (256) in each layer, as well as a network with 4 hidden layers and one with 6 hidden layers. The results of this regression are displayed in Fig. \ref{fig_net_arch}, and a quantitative comparison is given in Table \ref{tab_net_arch}. 

\begin{table}[h]
\caption{Error measures for different DNN architectures.}
\small
\centering
\begin{tabular}{|C{1.5cm}|c|C{1.5cm}|C{1.5cm}|C{1.5cm}|}
\cline{3-5}
\multicolumn{1}{c}{} & & 256 Neurons & 4 layers & 6 layers \\ \hline
\multirow{2}{*}{\begin{tabular}[c]{@{}c@{}}225 \\ Cross Sec.\end{tabular}}
& MSE & 0.0063 & 0.0046 & 0.0026 \\ 
& MPE & 5.121e-7 & 6.445e-8 & 5.388e-8  \\ \hline
\multirow{2}{*}{\begin{tabular}[c]{@{}c@{}}625 \\ Cross Sec.\end{tabular}}
& MSE & 0.0076 & 0.0045 & 0.0054  \\ 
& MPE & 9.025e-7 & 1.702e-7 & 5.857e-7  \\ \hline
\end{tabular}
\label{tab_net_arch}
\end{table}

The network architecture seems to play a larger role in regression quality than data sparsity. The network with 256 neurons suffers obvious visual degradation compared to the ground truth result. We also trained a network with 1024 neurons in each layer, however it failed to converge after 50 epochs. The network with 4 layers provides a better regression result, however, there are some visual artifacts visible in the IMF obtained with 625 cross sections. Finally, the 6 layer network provides a good regression result for the IMF with 225 cross-sections, however, it suffers degradation when generating the IMF with 625 cross sections (possibly from overfitting).

\clearpage

\end{document}